\documentclass[aps,preprint,tighten,floats,epsf,rotate]{revtex4}
\usepackage{array}
\usepackage{graphicx}
\usepackage{float}

\begin{document}
\title {Probing Dynamics of Phase Transitions occurring inside a Pulsar}
\author{Partha Bagchi$^1$ \footnote {email: partha@iopb.res.in}}
\author{Arpan Das$^1$ \footnote {email: arpan@iopb.res.in}}
\author{Biswanath Layek$^2$ \footnote{email: layek@pilani.bits-pilani.ac.in}}
\author{Ajit M. Srivastava$^1$ \footnote{email: ajit@iopb.res.in}}
\affiliation{$^1$ Institute of Physics, Sachivalaya Marg, 
Bhubaneswar 751005, India, \\ $^2$ Department of Physics, Birla Institute 
of Technology and Science, Pilani - 333031, India}

\begin{abstract}
\noindent During the evolution of a pulsar, various phase transitions
may occur in its dense interior, such as superfluid transition, as well as
transition to various exotic phases of quantum chromodynamics (QCD). We 
propose a technique which allows to probe these phases and 
associated transitions by detecting changes in rotation of
the star arising from density changes and fluctuations during the
transition affecting star's moment of inertia. Our results suggest 
that these changes may be observable, and may possibly account for 
glitches and (recently observed) anti-glitches. 
Accurate measurements of pulsar timing and intensity modulations
(arising from wobbling of star due to development of the off-diagonal
components of moment of inertia) may be used to pin down the particular 
phase transition occurring inside the pulsar core. We also discuss the 
possibility of observing gravitational waves from the changes in the
quadrupole moment arising from these rapidly evolving density 
fluctuations.\\

\end{abstract}

\maketitle

\section{introduction}
\noindent Exotic phases of quantum chromo dynamics (QCD), viz., 
quark-gluon plasma (QGP), color flavor locked (CFL) phase,
\cite{raja} etc. are possible at very high baryon density. The core of an 
astrophysical compact dense object, such as a neutron star, 
provides physical conditions where transition to these phases may be 
possible. Superfluid phases of neutrons (as well as of protons) are also
believed to exist inside neutron stars. Also, relatively young pulsars show 
the phenomenon of glitches \cite{glitch} and, recently
observed antiglitches \cite{antiglitch}, where the spinning rate of the 
pulsar rapidly changes, and then slowly relaxes. Conventional understanding 
of a glitch in terms of de-pinning  of a cluster of superfluid vortices in 
the core of a neutron star (which transfers angular momentum to the crust)
does not seem viable for explaining anti-glitch, though external
body impact has been proposed as a possible cause for antiglitches. 

In this paper we propose a technique to probe the dynamical phenomena 
happening inside the neutron star, which seems to be capable of also
accounting for the phenomena of glitches and anti glitches in a unified
framework. Basic physics of our approach is based on the fact that phase 
transitions are typically associated with density changes as well as
density fluctuation. Density fluctuation in the core of a star will in 
general lead to transient changes in its moment of inertia (MI), along with
a permanent change in MI due to phase transformation. This will directly 
affect its rotation and hence the pulsar timings. As accuracy of 
measurement of pulsar timings is extremely high 
($ \frac{\triangle \nu}{\nu}\sim 10^{-9}$), 
very minute changes of moment of inertia of star may be observable,
providing a sensitive probe for phase transitions in these objects. 
Non-zero off-diagonal components of moment of inertia arising from density
fluctuations imply that a spinning neutron star will develop wobble
leading to modulation of the peak intensity of pulses (as the direction
of the beam pointing towards earth undergoes additional modulation). 
This is a unique, falsifiable, prediction of our model, that 
rapid changes in pulsar timings should, most often, be associated with 
modulations in changes in peak pulse intensity. It is important to
note that the vortex de-pinning model of glitches is not expected to
lead to additional wobble as the change in rotation caused by de-pinning
of vortex clusters remains along the rotation axis. Density fluctuations will
also lead to development of rapidly changing quadrupole moment which can
provide a new source for gravitational wave emission due to extremely
short time scales involved (despite small magnitude of this new
contribution to the quadrupole moment) .

  The effect of phase changes on the moment of inertia has been discussed
in literature \cite{michange}. For example, moment of inertia change arising
from a phase change to high density QCD phase (such as to the QGP phase) is 
discussed in ref.\cite{michange}. The transition is driven by slow decrease
in rotation speed of the pulsar, leading to increasing central density
causing the transition as the central density becomes supercritical. 
It is assumed that as the supercritical core grows in size (slowly,
over the time scale of millions of years), it continuously converts
to the high density QGP phase (even when the transition is of first order). 
Due to very large time scale, the changes
in moment of inertia are not directly observable, but observations of
changes in the braking index may be possible.

   Our work differs from these earlier discussions in two important
aspects. We emphasize the possibility that the transition to the high
density (say QGP) phase may not happen continuously. This will happen
when the (first order) transition is strong, with relatively large
latent heat. In such a situation, the transition will happen by 
nucleation of bubbles of high density phase in the superdense core.
The superdense core may become macroscopically large (of sizes meters 
to even Km) before a single bubble may nucleate. (This is exactly how
large nucleation distances, of the order of meters, could be
possible in the original discussion of Witten \cite{witten} in
the context of quark hadron transition in the early universe.) Once a bubble 
is nucleated, it will expand with large speed (possibly relativistically, 
with the speed of sound) converting the entire supercritical core to the 
QGP phase (assuming, as is usually done, fast dissipation of the latent 
heat). Thus in this scenario, though the supercritical core may grow to
macroscopic size over the time scale of millions of years, its
conversion to the high density phase may occur in very short time, say, 
in micro seconds. (Such rapid transitions have only been discussed in 
the context of {\it hot} neutron stars during its very early stages 
\cite{bblnucl} where transition proceeds by thermal nucleation of bubbles.) 
The associated changes in the moment of inertia,
hence the pulsar timing, will be on very short time scales and should be 
directly observable. In fact, we will argue below that some of the glitches 
and antiglitches could be due to such rapid phase transitions. 

   The second aspect in which our work differs from the earlier works lies
in our focus on the density fluctuations arising during the phase transitions.
This has not been considered before as far as we are aware. Density 
fluctuations inevitably arise during phase transitions, e.g. during
a first order transition in the form of nucleated bubbles, and may
become very important in the critical regime during a continuous 
transition. The density fluctuations arising from a phase transition
become especially prominent if the transition leads to formation of 
topological defects. Extended topological defects can lead to strong 
density  fluctuations which can last for a relatively long time (compared 
to the phase transition time). It is obvious that such randomly arising
density fluctuations will affect the moment of inertia of the star
in important ways. Most importantly, it will lead to development 
of transient non-zero off-diagonal components of the moment of inertia,
as well as transient quadrupole moment. Both of these will disappear after
the density fluctuations decay away and the transition to a uniform
new phase is complete. Net change in moment of inertia will have this
transient part as well as the final value due to change to the new phase.
It seems clear that this is precisely the pattern of a glitch or 
anti-glitch where rapid change in pulsar rotation is seen which slowly 
and {\it only partially} recovers to the original value. Transient change in
quadrupole moment will be important for gravitation wave emission, due
to extremely short time scale associated with the evolution of these
density fluctuations, as we will explain below.
 
\section{Change in moment of inertia due to a first order transition} 

As we mentioned above, the discussions in ref.\cite{michange} about
the change in moment of inertia due to a phase transition assumed
that the phase conversion happens continuously in the supercritical 
region of the core. This will happen for a second order transition,
or a crossover, or for a weak first order transition with very large
bubble nucleation rate. However, for a strong first order phase transition,
this may not happen. As we mentioned above, for very low nucleation
rates, the supercritical core may become macroscopically large before
a single bubble of new phase nucleates. Once nucleated, the bubble
will expand fast sweeping entire supercritical core and converting it
to the new phase. This will lead to the change in the moment of inertia
of the pulsar in a very short time which may be directly observable.
A rough estimate of change in the moment of inertia due to phase change 
can be taken from ref.\cite{michange} using Newtonian approximation, and
with the approximation of two density structure of the pulsar. If the density
of the star changes from $\rho_1$ to a higher density $\rho_2$ inside
a core of radius $R_0$, then the fractional change in the moment of
inertia is of order,

\begin{equation}
{\Delta I \over I} \simeq {5 \over 3} 
({\rho_2 \over \rho_1} - 1){R_0^3 \over R^3}
\end{equation}

Here $R$ is the radius of the star in 
the absence of the dense core. For a QCD phase transition, density changes
can be of order one. We take the density change to be about 30\% as an
example. If we take the largest rapid fractional change in the moment  of 
inertia of neutron stars, observed so far (from glitches), to be less than 
$10^{-5}$, then Eq.(1) implies that $R_0 \le$ 0.3 Km (taking $R$
to be 10 Km). For a superfluid transition, we may take
change in density to be of order of superfluid condensation energy
density $\simeq$ 0.1 MeV/$fm^3$ (see, ref.\cite{superfluid}).
In such a case, $R_0$ may be as large as 5 Km. These constraints on
$R_0$ arise from observed data on glitches/anti-glitches. These estimates
may also be taken as prediction of possible large fractional changes in 
the moment of inertia (hence pulsar spinning rate) of order few percent
when a larger core undergoes rapid phase transition. For example, $R_0$
may be of order 2-3 km for QCD transition (from estimates of high density 
core of neutron star \cite{nstar}), or it may be only slightly smaller 
than $R$ for superfluid transition.

   Let us now discuss the conditions which will allow such a rapid phase
transition in  a large core. As an example, we consider a simple case
of zero temperature transition (as appropriate for late stages of
neutron star) between a nucleonic phase and a QGP phase with pressures
($P$) and energy densities ($\epsilon$) of the two phases given as 
follows \cite{qgp}. 

\begin{equation}
P_{nucleon} = {M^4 \over 6\pi^2} \left ( {\mu \over M}
\left({\mu^2 \over M^2} - 1 \right)^{1/2} \left({\mu^2
\over M^2} - {5 \over 2} \right)
+ {3 \over 2} 
ln \left[{\mu \over M} + \left({\mu^2 \over M^2} -1 
\right)^{1/2} \right] \right)
\end{equation}    

\begin{equation}
\epsilon_{nucleon} = {2\mu \over 3\pi^2} (\mu^2 - M^2)^{3/2}
-P_{nucleon}
\end{equation}

\begin{equation}
P_{QGP} = {\mu_q^4 \over 2\pi^2} - B
\end{equation}

\begin{equation}
\epsilon_{QGP} = 3P_{QGP} + 4 B
\end{equation}

Here $\mu_q$ and $\mu (= 3\mu_q)$ are the baryon chemical
potentials for quarks and nucleons respectively, $M$ is the mass of
the relevant hadron (nucleon), and $B$ is the bag pressure. Note that 
with this simple {\it Bag model} equation of state, the transition to
QGP phase requires absorption of latent heat which should be
provided by the release of gravitational potential energy from
the compression of the core. For other equations of state
(see, e.g. \cite{bblnucl}), or for other QCD transitions (say from
QGP to CFL phase) the transition may release latent heat
which will be rapidly dissipated by the star.

  The nucleation rate appropriate for zero (low) temperature is
dominated by quantum tunneling mediated by O(4) symmetric
instantons, and is given by \cite{clmn}.

\begin{equation}
\Gamma = A {S_0^2 \over 4\pi^2} exp(-S_0)
\end{equation} 

where $A$ is the determinant of fluctuations around the instanton 
configuration, and $S_0$ is the Euclidean action of the instanton.
In our case, we will be interested in the situation of extremely low nucleation
rates, corresponding to very large values of action $S_0$. The nucleation
rate then will be completely dominated by the exponential factor, and
the pre-exponential factor can be approximated by dimensional estimates using
$A = R_c^{-4}$ where $R_c$ is the radius of the critical bubble (size
of the instanton in Minkowski space). Recall, that at finite temperature
$T$, dimensional estimates use $A = T^4$. The action $S_0$ for the instanton
can be obtained from the action for an O(4) symmetric configuration 
written as follows,

\begin{equation}
S = -{1 \over 2} \pi^2 R^4 \Delta P + 2 \pi^2 R^3 S_1
\end{equation}

where $\Delta P$ is the pressure difference between the two phases and
$S_1$ is the action of a one-dimensional instanton giving the contribution
of the surface term of the bubble to the action. Extremization of
$S$ gives the critical radius $R_c = 3S_1/(\Delta P)$  with which
the action of the instanton $S_0$ is found to be

\begin{equation}
S_0 = {27 \pi^2 S_1^4 \over 2 (\Delta P)^3}
\end{equation}

 For our case, $\Delta P = P_{QGP} - P_{nucleon}$ (Eqns. (2),(4)). For
calculation of $S_1$, one needs the free energy functional (e.g.
Landau-Ginzburg free energy). In the absence of that we simply
consider a range of values of surface tension $S_1$ ranging from
0.01 MeV/$fm^2$ to 5 MeV/$fm^2$. As we discussed above, for QCD
scale phase transitions, observations constrain the critical core 
size to be less than about 0.3 Km. We calculate number 
of bubbles nucleated in 300 meter radius core in one million
year time duration as a function of core density. This is
given in Fig.1a. For this we have used parameters $B^{1/4}$ = 
177.9 MeV, surface tension $S_1 \equiv \sigma = $ 0.05 MeV/$fm^2$
and $M$ = 1087.0 MeV (taken as the mean of the nucleon and delta mass, 
ref. \cite{qgp}). Value of surface tension is unusually small here.
With the simple equations of state for the two phases used
here, nucleation rate rapidly drops with much larger values of
$\sigma$. For a more realistic equation of state, larger values of
surface tension may be possible. (Note that we are considering homogeneous
nucleation here. There may be inhomogeneities in the core region
enhancing the nucleation probability via hetrogeneous nucleation.)
With these choices, the critical density for the transition is
found to be $\rho_c = 2.500 \rho_0$ (where $\rho_0 \simeq 0.15 m_{nucleon}$ 
is the nuclear saturation density). Fig.1 shows that at a density 
$\rho_{nucl} \simeq  2.502 \rho_0$ the number of nucleated bubble 
is one. The critical radius of the bubble $R_c = 50$ fm at this density.
Nucleation rate changes sharply as a function of density, and is 
insignificant at lower values of $\rho$. For example, with a decrease in
density by only 0.01\%, the number of bubbles nucleated is about $10^{-10}$.
 
\begin{figure}[ht!]
\includegraphics[width=12cm]{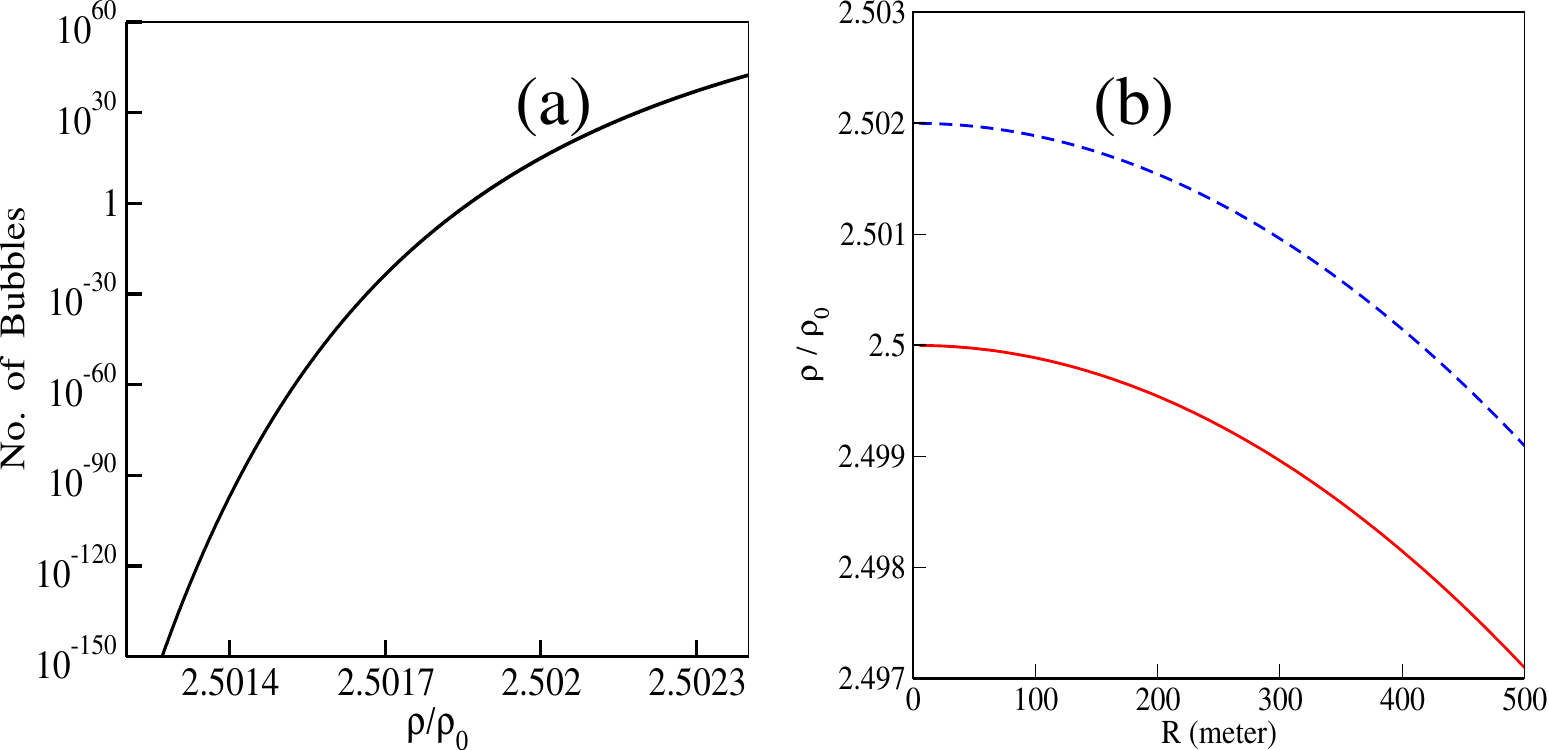}
\leavevmode
\caption{(a) Plot of the number of bubbles nucleated in 300 meter radius 
core in one million year time duration as a function of core density
for a QCD transition. (b) Solid plot shows the density profile of the core 
region of the neutron star with neutron star mass is $M_1 = 1.564 M_0$. 
Density at $r = 0$ has just reached the critical value $rho_c$. The dashed 
plot shows the density profile when the supercritical core 
size (with $\rho > \rho_c$) has increased to about 300 meter. The mass 
of the neutron star at this stage is $M_2 = 1.567 M_0$.}

\end{figure}
\vspace{0.25cm}

  For density change of the neutron star, we consider the case of
accretion driven change. Typical data shows that neutron stars
in a binary system accrete matter at the rate of about $10^{15} - 
10^{18}$ grams/sec \cite{acrete}. With accretion rate of $10^{17}$
grams/sec, for a solar mass neutron star, this will 
mean about 0.1\% change in its mass in one million year. We calculate density
profile of a non-rotating star (this approximation will be valid for
slowly rotating pulsars) in Newtonian approximation using a polytropic
equation of state $P = K \rho^\alpha$. We take $\alpha = 2.54$ with 
$K = 0.021 \rho_0^{-1.54}$ (as in ref.\cite{michange})
and solve the following equation for 
density profile \cite{michange}. 

\begin{equation}
{1 \over \rho}{dP \over dr} = -{Gm \over r^2} ~; ~~ dm = 4\pi r^2 \rho dr
\end{equation}

With central density $\rho = 2.5 \rho_0$ (= $\rho_c$, the critical density
for hadron-QGP transition) we get neutron star mass to be 1.564 $M_0$ 
($M_0$ is the solar mass) and its radius to be about 13.7 Km. We consider
the situation when a neutron star with sub-critical central density
accretes matter such that its central density becomes super-critical.  
Consider the stage when the central value of the density of the 
neutron star (at radius $r = 0$) just reaches the critical density $\rho_c$
by mass accretion. We calculate the density profile with $\rho = \rho_c$
at the center $r = 0$. This situation is shown by the (solid) plot in 
Fig.1b showing the density profile of the core region of the neutron star.  
Mass of the star at this stage (with our choice of parameters) is
$M_1 = 1.564 M_0$.  Subsequently, the continued accretion increases
the size of the core region where the density is supercritical
($\rho > \rho_c$).  The dashed plot in Fig.1b shows the density profile
of the core region when the supercritical core size has increased
to about 400 meter. The mass of the neutron star at this stage is
$M_2 = 1.567 M_0$. Taking the accretion rate of $10^{17}$ grams/sec.
it will take about one million year for the supercritical core
size to increase to this size. Nucleation rate in Fig.1a shows that
at this stage there can be just about one bubble nucleation
possible in this supercritical core. (Actually core needs to be
somewhat larger as in regions away from the center in the supercritical
core, nucleation rate is smaller. This difference is unimportant
for our rough estimates).

   Once the bubble is nucleated, it will sweep through the entire
supercritical core. Typical speed of bubble wall propagation will
be relativistic, and one may take the speed of sound in a relativistic
plasma (c/$\sqrt{3}$) as an estimate. Thus the transition will be
completed in time of order microsecond. The resulting transition is
therefore completed in a very short time, even though the supercritical
core grew over a million year time. We again point out that this scenario
is very similar to the one discussed by Witten \cite{witten} for the
early universe where bubble nucleation is insignificant until the
age of the universe is few microseconds (many orders of magnitude
larger than the strong interaction time scale), and inter-bubble
separation of order centimeters or even meters is possible.
As discussed above, such a rapid transition in a macroscopically
large core of the neutron star  will lead
to fractional change in moment of inertia of order $10^{-5} - 10^{-6}$
which may be observed  as a pulsar glitch. The supercritical core 
size (in which a single bubble can nucleate) may be  larger (for
appropriate values of parameters such as bubble surface tension).
In that case fractional change in moment of inertia of even few percent
(occurring in a very short time of order micro seconds) may be possible 
and this can be taken as a prediction of our model suggesting lookout
for such candidates. 

  One more consequence of the scenario discussed above has 
implication for the superfluid phase of the neutron star. We consider
the above discussed transition such that latent heat of order few hundred 
MeV/$fm^3$ is released in the QCD scale transition.
Assuming that most of the neutron star is in the neutron superfluid
phase (and/or proton superconducting  phase) at this stage, with the 
free energy density
scale for the superfluid transition being of order 0.1 MeV/$fm^3$, the
latent heat released by the QCD transition will heat up the superfluid
phase to the normal phase. Simple volume ratio will tell that the latent
heat released in a 300 meter core undergoing QCD transition will convert
about 3 km radius region from superfluid to the normal phase as the heat 
pulse sweeps through the neutron star. Subsequent cooling will again lead
to transition to the superfluid phase for all that region. Our estimates
of change in moment of inertia above suggest that even this superfluid 
transition happening in radius of about 3 km will again lead to fractional
change in moment of inertia of order $10^{-5} - 10^{-6}$. Again, a larger
core will lead to larger change in moment of inertia.

\section{Density fluctuations due to bubble nucleation}

 We now focus on the effects of density fluctuations generated during
the phase transition on the moment of inertia and the quadrupole
moment of the neutron star.

The core of a neutron star may go through a transition from hadronic 
matter to QGP due to gradual slowing down of the rotating neutron star
\cite{michange}, or due to accretion of matter. 
Or a QGP core formed during early hot and dense phase of
neutron star may undergo a transition to hadronic matter after a 
relatively longer time as the core cools. At these baryon densities,
the transition is very likely a first order transition and we first 
focus on density fluctuations generated due to nucleation of
bubbles. As we discussed above, for strong first order case, a core
of size few hundred meter (or larger) can become supercritical without
any bubble nucleation taking place inside the core. With further accretion
of matter the density may become {\it sufficiently} large so
that bubble nucleation can take place. We discussed the case above when
a single bubble nucleation takes place. Consider now the situation when
density change is such that no bubbles are nucleated until density
increases to a value when a reasonably large number of bubbles (several
thousand) can nucleate inside the supercritical core. With strictly ideal,
monotoic decrease in density with radial distance such a scenario looks
unlikely. However, we emphasize that in general the core region will
be expected to have minute nonuniformities, even of purely statistical
origin. For example, the temperature of different parts of the core
(even at same radial distance, but in different directions) cannot
have exactly the same value. Purely from statistical fluctuations,
there will be fluctuations in temperature in these regions (similarly
in chemical potential) which will depend on properties such as specific
heat \cite{landau}. In fact, such density fluctuations can lead to
large enhancement in our estimates of density fluctuations from
defect formations. This is because density of defects is entirely 
determined by the correlation length which sensitively depends
on parameters like temperature, chemical potential etc. Varying
correlation length will lead to additional source of
fluctuation in the density of defects, hence for density fluctuations.
We will not get into such details here, but only
conclude that a situation where many bubbles may nucleate in different
parts of supercritical core may not be unreasonable for a realistic
case.  After nucleation, bubbles (with initial critical size being
microscopic, of order tens of fm) rapidly expand and
coalesce. At the time of coalescence, the supercitical core region
will consist of a close packing of bubbles of new phase, embedded in
the old phase. We assume that the latent heat is released from the star
which either contributes to a uniform background in energy density
(contributing to the net moment of inertia of the star as a homogeneous
sphere), or it is simply dissipated away from the star. In either case,
the latent heat will not affect the off-diagonal component of the moment
of inertia  and the quadrupole moment.   

  We simulated random nucleation of spherical bubbles of radius $r_0$ 
(at the coalescence stage) 
filling up a spherical core of radius $R_0$, with density change
of order few hundred MeV/$fm^3$ (as appropriate for a QCD scale transition.
For $R_0 \simeq 300 $meters we find fractional change in moment of
inertia $\Delta I/I \simeq 4 \times 10^{-8}$ for $r_0 = 20$ meters. Change in
moment of inertia remains of same order when $r_0$ is changed from 20 meter
to 5 meter. Due to random nature of bubble nucleation, off-diagonal
components of the moment of inertia, as well as the quadrupole moment
become nonzero and the ratio of both to the initial moment of inertia 
are found to be of order $10^{-11} - 10^{-10}$. 
This aspect of our model is extremely
important, arising entirely due to density fluctuations generated during
the transition. As these density fluctuations homogenize, finally leading
to a uniform new phase of the core, both these components will dissipate 
away. The off-diagonal component of moment of inertia will necessarily
lead to wobbling (on top of any present initially), which will get
restored once the density fluctuations die away. This will lead to
transient change not only in the pulse timing, but also in the pulse
intensity (as the angle at which the beam points towards earth gets
affected due to wobbling).  We again emphasize that 
the conventional vortex de-pinning model of glitches is not expected to
lead to additional wobble as the change in rotation caused by de-pinning
of vortex clusters remains along the rotation axis. 

    Generation of quadrupole moment has obvious implication for gravitational
wave generation. One may think that a quadrupole moment of order $10^{-10}$
is too small for any significant gravity wave emission. However,
note that the gravitation power depends on the (square of) third time 
derivative of the quadrupole moment \cite{gwave}. The time scales
will be extremely short here compared to the time scales considered in
literature for the usual mechanisms of change in quadrupole moment
of the neutron star. Here, phase transition dynamics will lead to
changes in density fluctuations occurring in time scales of microseconds
(or even shorter as we will see below in discussions of topological defects
generated density fluctuations). This may more than compensate for
the small amplitude of quadrupole moment and may lead to these density
fluctuations as an important source of gravitational wave emission from
neutron stars.

\section{Density fluctuations from topological defects}

 Formation of topological defects in symmetry breaking transitions has
been extensively discussed in the literature, from the early universe
to condensed matter systems. Topological defects form during spontaneous 
symmetry breaking transitions via the so called 
{\it Kibble mechanism} \cite{kbl}. These defects can be source of
large density fluctuations depending on the relevant energy scales, and
we now focus on these density fluctuations. The defect network resulting
from a phase transition and its evolution shows universal 
characteristics, e.g. defects have initial
densities which basically depend only on the correlation length and on the
relevant symmetries and the evolution of string defects and domain wall 
defects shows scaling behavior. This has important implications for our model
as  the universal properties of
defect network and scaling during evolution may lead to reasonably
model independent predictions for changes in moment of inertia (hence
glitches/anti-glitches) and quadrupole moment, and subsequent relaxation to 
original state of rotation.

  Most important aspect of these changes in moment of inertia components
(especially the off-diagonal ones) and the quadrupole moment arising
from density fluctuations during phase transition is the following. Specific
pattern of density fluctuation, and the manner in which it decays (to
eventual uniform new phase) crucially depends on the nature of the source
of the fluctuations. Bubbles, strings, domain walls, all generate different
density fluctuations, and detailed simulations can determine the nature
of resulting changes in pulsar timings and intensities (due to wobbling
as discussed above) resulting from these. High precision measurements
have potential of distinguishing between different sources of fluctuations,
thereby pinning down the specific phase transition occurring inside the
core of a neutron star. Similarly, the evolution of density fluctuation 
also depends on the specific case being considered. For example, bubble 
generated density fluctuations discussed above will decay away quickly in 
time scale of coalescence of bubbles, while domain wall network and the
string network may coarsen on much larger time scales ( and differently 
from each other). Thus the relaxation of the pulsar spinning (and
wobbling etc.) can also provide important information about the
specific transition (leading to corresponding defect formation) occurring 
inside the neutron star.

\subsection{Field theory simulations  for QCD transition}

 First we consider transition between various exotic high density
QCD phases. Phase transition from hadronic phase to QGP phase, and 
further, from QGP phase to the CFL phase etc. are also possible due to 
a slow evolution of core density, either by accretion of matter from a 
companion star, or due to slow decrease in rotation velocity, all these effects
lead to increase of baryon density in the core. We will not worry 
about the detailed cause of a phase transition, and simply assume that a 
phase transition occurs in the core of the neutron star. We now consider
hadron-QGP phase transition, i.e. confinement-deconfinement (C-D) phase
transition. The expectation value of the Polyakov loop, $l(x)$, is the 
order parameter for this transition \cite{plkv}. $l(x)$ is zero in the 
hadronic phase and is non-zero in the QGP phase breaking Z(3) center 
symmetry (for the SU(3) color group) spontaneously
as $l(x)$ transforms non-trivially under $Z(3)$. This gives rise to 
topological domain wall defects in the QGP phase which interpolate 
between different $Z(3)$ vacua and also string defects (QGP strings) 
forming at the junction of these $Z(3)$ walls \cite{zn,layek1,ams}.
We point out that in the presence of quarks, $Z(3)$ symmetry
is also explicitly broken and it affects the dynamics (especially
at late times) of $Z(3)$ walls
and the QGP string in important manner \cite{ams}. However, again
for an order of magnitude estimate, we confine our focus on very early
stages of evolution of defect network and neglect these quark effects.
We carry out a field theory simulation of the evolution of $l(x)$
from an initial value of zero (appropriate for the hadronic phase) as
the system is assumed to undergo a rapid transition (quench) to the QGP
phase (as in \cite{quench}). Use of quench is not an important point 
here as the formation of
defects only requires formation of uncorrelated domains, and the size of
the domains in our model has to be treated as a parameter. 

 We mention that the estimates of change in moment of inertia due to the 
formation of the QGP string and $Z(3)$ domain walls etc. require 
microphysics governed by QCD scale of order $10^{-15}$ meters, while the 
star radius is in km.  It is not possible for
us to carry out simulation covering such widely different scales of
length and time. Results with appropriate length scales, as for the 
bubble nucleation case, (Sect.III) are not possible for general case and
one has to resort to simulations. We now discuss detailed
field theory simulations, which are necessarily restricted to
very small system sizes. For the simulation we use, as an example, 
the effective Lagrangian proposed in \cite{psrsk}. 

\begin{equation}
L = \frac{N}{g^2} |\partial_\mu l|^2 T^2 - V(l) .
\end{equation}
Here, $N = 3$ and $V(l)$ is the effective potential for the Polyakov loop 
given by,
\begin{equation}
V(l) = \Bigl( -\frac{b_2}{2}|l|^2 - \frac{b_3}{6}\bigl(l^3 + (l^*)^3 
\bigr) + \frac{1}{4} \bigl(|l|^2 \bigr)^2 \Bigr) b_4 T^4 .
\end{equation}
$l_0$ is given by the absolute minimum of $V(l)$ and the normalization of 
$l(x)$ is chosen such that $l_0 \rightarrow 1$ as $T \rightarrow \infty$.  
Values of various parameters in Eq.(4) are fixed with lattice result 
following Ref.\cite{psrsk} (see refs. \cite{layek1,ams}
for these details). Time evolution of $l(x)$ is governed
by the field equations obtained from Lagrangian in Eq.(10).
We use leap frog algorithm with periodic 
boundary conditions for the simulation. The physical size of the lattice 
is taken as (7.5 fm)$^3$ and (15 fm)$^3$ with lattice spacing, 
$\Delta x$ = 0.025 fm and 
time step, $\Delta t = \frac{0.9 \times \Delta x}{\sqrt{3}}$. To minimize 
the effects of periodic boundary conditions, a spherical region
with radius $R_c$ is chosen to study change of moment of inertia, with
$R_c$ = 0.4(lattice size). This represents the core of the neutron star. 
We use temperature T = 400 MeV as a sample value. Note that temperature of
few hundred MeV is needed to get  correct energy scale of QCD transition
for field theory model of Eqn.(10) which corresponds to lattice data
at zero chemical potential. This value of $T$ has nothing to do with the 
actual temperature of the neutron star where QCD transition can occur at 
very small temperatures due to high baryon density. We add a dissipation term 
to enable relaxation of density fluctuations so that finally homogeneous
phase is reached completing the phase transition. The decrease in
energy due to dissipation term is added as a uniform background to
the core energy to keep the total energy fixed. Note that with this
assumption we are ignoring the change in the moment of inertia due
to  net change in the phase of the core. The reason we are forced to 
do this is because there are numerical errors (of order few percent)
in evolving a field theory configuration via leapfrog algorithm.
Since we are looking for fractional changes of order $10^{-6}$
or even smaller, numerical errors will mask any such changes.
Thus we keep the net energy fixed and only focus on re-distribution
of energy in defect network and the background. For the net change
in the moment of inertia, we will use the estimates from Eqn.(1)
(\cite{michange}) as discussed in Sect.II.

We shall now discuss the third possibility of phase transition to the
so called color flavor locked (CFL) phase inside the core of a pulsar 
where the QCD symmetry for three massless flavors, $SU(3)_c \times 
SU(3)_L \times SU(3)_R \times U(1)_B$ is broken down to the diagonal 
subgroup $SU(3)_{c+L+R} \times Z_2$ at very high baryon density 
\cite{raja} by the formation of a condensate of quark Cooper pairs.
This transition will give rise to global strings (vortices). To roughly
estimate resulting change in MI, we consider a
simplified case by replacing the cubic term $(l^3 + {l^*}^3)$ in Eq.(11) 
by  $(|l|^2 + |l^*|^2)^{3/2}$ term.  This modification in the potential 
will give rise to string defects only without any domain walls, as 
appropriate for the transition from (say) QGP phase to the CFL phase, 
while ensuring that we have the correct energy scale for these string 
defects. 

\begin{figure}[ht!]
\includegraphics[width=17cm]{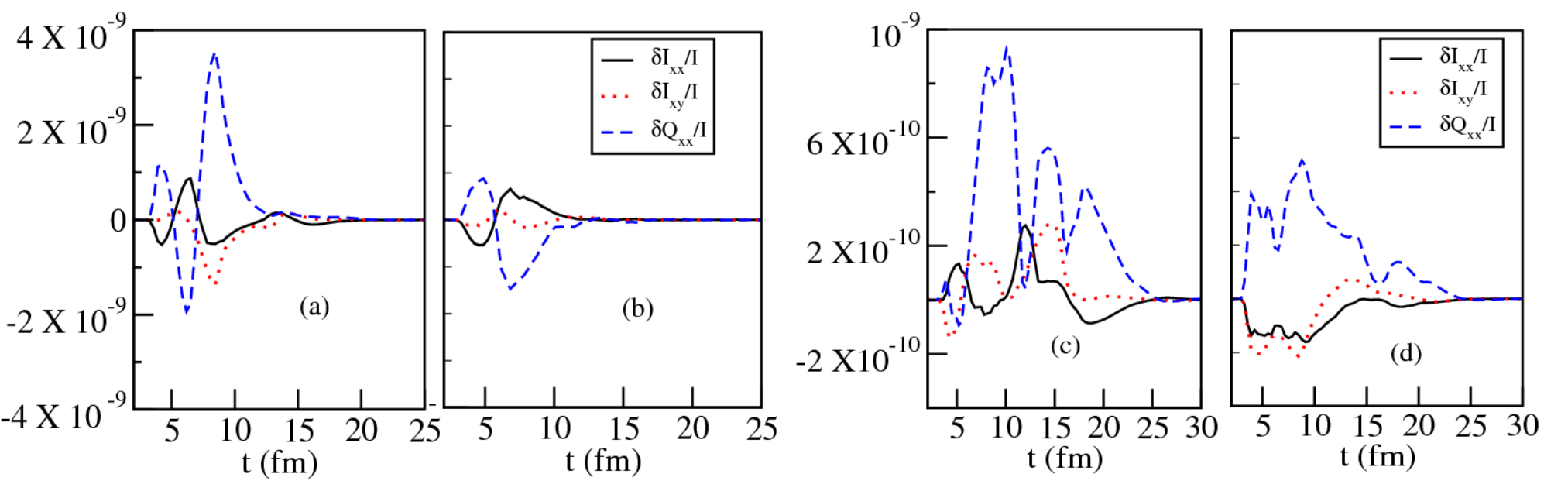}
\leavevmode
\caption{Fractional change in moment of inertia and quadrupole moment 
during phase transitions. (a),(b) correspond to 
lattice size (7.5 fm)$^3$, and (c),(d) correspond to lattice size 
(15 fm)$^3$ respectively. Plots in (a) and (c) correspond to the 
confinement-deconfinement phase transition with Z(3) walls and strings,
while plots in (b) and (d) correspond to the transition with
only string formation as appropriate for the CFL phase.} 
\end{figure}
\vspace{0.25cm}

The results of field theory simulations for C-D phase 
transition and the transition with only string formation as appropriate 
for the CFL phase are summarized in Fig.2. Here the plots show the time 
evolution of the fractional change in the MI of the core relative 
to the initial MI of the full star.  Here we have considered QCD transition
occurring in the dense core of fractional size  0.3/10. This is in
accordance with the discussion above that the change in MI of a 10 km 
neutron star will be less than about 10$^{-5}$ if the QCD scale
transition occurs inside a core of size smaller than about 300 meters.
As discussed above, we can also consider larger core sizes which will lead 
to change in MI of star, of order few percent, which has not been seen.
Nonetheless, this possibility remains a prediction of our model. 
For initial MI of the full star
we add the MI of a shell outside the core so that the star size = 
${10 \over 0.3}$ of the core size, and the shell has the same uniform  
density as the core.  The magnitudes of these fractional changes due to
density fluctuations for all the diagonal components, $I_{xx}, I_{yy}$ 
and $I_{zz}$ are found to be of same order, though the changes
for different components may be positive or negative. Note again
that here we are only presenting change in MI due to density fluctuations
during phase transition. Net change in MI will include the very
large contribution of order 10$^{-5}$ due to net phase change of the
core (\cite{michange}). The change due to fluctuations is the transient
one and will dissipate away as star core achieves uniform new phase.
As we see, the transient change have either sign, similarly the net
change can also have either sign depending on the nature of the transition
(QGP to CFL, or reverse transition, or hadronic to QGP etc.).
Thus, this evolution pattern of fractional changes in MI suggests that the 
phase transition dynamics may be able to account for both glitch and 
anti-glitch events. We have also observed the development of off-diagonal 
components, $I_{xy}, I_{xz} \& I_{yz}$ (Fig.2). The small change 
in these components will cause the pulsar to wobble about its axis of 
rotation. This will lead to modulation of the peak intensity of pulse. This
is a definitive, falsifiable, prediction of our model. Such phenomenon, if 
observed will be a signal for random density fluctuations occurring inside 
the star, strongly indicating a phase transition.  Changes in the quadrupole
moment will lead to gravitational wave emission and we will discuss it
further in the next sub-section. 

\subsection{String and wall simulation} 

 We also estimate change in MI due to string and wall formation by taking an 
alternative static approach. We produce a network of defects inside the 
core of the pulsar by modeling the correlation domain formation
in a cubic lattice, with lattice spacing $\xi$ representing the correlation
length \cite{tanmay}. Each lattice site is associated with an angle 
$\theta$ (randomly varying between 0 and 2$\pi$ (to model U(1) global 
string formation), or two discrete values 0,1 when modeling $Z_2$ domain 
wall formation. (For simplicity we consider $Z_2$ domain walls instead of
$Z_3$ walls of QCD). For string case, winding of $\theta$ on each face 
of the cube is determined using the geodesic rule \cite{kbl,tanmay}. 
For a non-zero winding, a string segment (of
length equal to the lattice spacing $\xi$ ) is assumed to pass through that
phase (normal to the phase). For domain wall case, any link connecting
two neighboring sites differing in $Z_2$ value is assumed to be intersected 
by a planar domain wall (of area $\xi^2$, and normal to the link). We consider 
spherical region inside the lattice with all lattice sites under consideration 
being within the radius of the core. This leads to a somewhat zigzag
boundary, but for lattice sizes larger than 100$^3$ this effect is negligible.
The mass density (i.e.  mass per unit length) of the string was taken as 
3 GeV/fm, and the domain wall tension is taken to be 7 GeV/fm$^2$, as 
was estimated by numerical minimization technique in 
Ref. \cite{layek1} for the pure gauge case at T = 400 MeV. (For lower 
temperature, this value will be somewhat lower, which will reduce our 
results for change in MI by a similar factor.) The mass of the star core
is reduced by the amount of energy-mass contained in the defect network.
It is important to note here that for the transitions considered here, the 
string defect corresponds to global symmetry. Global strings have much 
larger energy densities associated with them (with logarithmic dependence on 
inter-string separation). A proper account of this can lead to much
larger density fluctuations than considered here.

We consider spherical star of size $R$ and confine
defect network within a spherical core of radius $R_c = 
{0.3 \over 10} R$.  For $\xi \simeq $ 10 fm, we find the resulting value 
of  $\frac{\delta I}{I_i} \simeq 10^{-12} - 10^{-13}$ implying
similar changes in the rotational frequency. Here $I_i, i = 1,2,3$ are
the three diagonal values of the moment of inertia tensor. As we increase
$R_c$ from 5 $\xi$ to about 400 $\xi$, we find that the value of
$\frac{\delta I}{I_i}$ decreases first, and then stabilizes near
$ 10^{-13} - 10^{-14}$ as shown in the table below (somewhat larger values 
are seen for small $R_c$ due to relatively larger fluctuations). 
This range of values of $R_c$ amounts to change in the number of  string 
and wall segments by $10^6$. This gives a strong possibility
that the same fractional change in the moment of inertia may also be
possible when $R$ is taken to have the realistic value of about 10 Km,
supporting the validity of extrapolation assumed in this paper. For
the formation of domain walls (again, with domain size of order 10 fm)
we find fractional change in moment of inertia components (as well
as quadrupole moments) to be larger by about a factor of 40, i.e.
of order $10^{-12}$. We should re-emphasize that with the possibility
of statistical fluctuations in temperature, chemical potential etc.
inside the core, density fluctuations due to defects can significantly
increase leading to much larger changes in MI and quadrupole moment.

  This change in moment of inertia is very small, and possibly difficult
to observe at the present stage. With increase in precision, it
should be observable. Also, if the core size undergoing transition
is taken to be larger (with resulting change in the moment of inertia
due to phase change larger than $10^{-5} - 10^{-6}$) then defect
induced change can also be larger.

\begin{table}[h!]\footnotesize
\caption{Fractional change of various components of moment of inertia
and quadrupole moment  caused by 
inhomogeneities due to random distribution of string network in the core 
of the pulsar. The results are obtained by varying core size, $R_c$ while 
keeping the correlation length $\xi =$ 10 fm fixed.}
\vspace{0.25cm}
\begin{tabular}{|c|c|c|c|c|c|c|c|c|c|}
\hline
& \multicolumn{3}{|c|}{QCD Strings} & \multicolumn{3}{|c|}{QCD Walls} & \multicolumn{3}{|c|}{Superfluid Strings}\\
\hline
$\frac{R_c}{\xi}$ & $\frac{\delta I_{xx}}{I}$  & $\frac{\delta I_{xy}}{I}$
& $\frac{Q_{xx}}{I}$ &$\frac {\delta I_{xx}}{I}$ & $\frac {\delta I_{xy}}{I}$ & $\frac {Q_{xx}}{I}$
&$\frac {\delta I_{xx}}{I}$ & $\frac{\delta I_{xy}}{I}$ &$\frac {\delta Q_{xx}}{I}$\\
\hline 
5 & 5.0E-10 & -3.0E-10 & -1.3E-10 & 1.9E-8 & -9.6E-9 & -7.7E-10 &1.7E-6 &-9.9E-7 &-4.3E-7\\
\hline 
50 & 5.1E-12 & -2.2E-12 & 2.1E-12 &1.4E-10 & -7.5E-11 & -1.2E-11 &1.7E-8 &-7.2E-9 &7.0E-9\\
\hline
100 & 1.0E-12 & -8.4E-13 & -3.8E-13 & 4.5E-11 & -2.4E-11 & 8.5E-12 & 3.4E-9 & -2.8E-9 & -1.3E-9 \\
\hline
200 & 1.4E-13 & 1.7E-14 & -6.7E-14 & 5.0E-12 & -4.3E-12 & -5.8E-12 & 4.8E-10 & 5.7E-11 & -2.2E-10 \\
\hline
300 & 1.8E-13  & -9.6E-15 & 1.2E-13 & 2.8E-12 & -2.1E-13 & -5.1E-12 & 5.9E-10 & -3.2E-11 & 4.0E10 \\
\hline
400 & -3.3E-15 & -5.3E-14 & -9.1E-14 & 2.7E-12 & -2.1E-12 & 3.4E-14 & -1.1E-11 & -1.8E-10 & -3.0E-10\\
\hline
\end{tabular}
\end{table}

 As discussed above, an important prediction of our model is generation of 
non-zero quadrupole moment of the star. Table I also presents the typical 
value of the ratio of quadrupole moment to the moment of inertia of the 
star for our random defect formation model, as star core size is increased 
from 5 $\xi$ to 400 $\xi$. We note that this ratio stabilizes (within about
an order of magnitude) about a value $\simeq 10^{-13}$. 
Even if this magnitude turns out to be much smaller than the quadrupole 
moment due to deformation of the star, the power emitted in
gravitational waves may not be small due to very short time scales
in the present case. The string coarsening
will be governed by microphysics with QCD time scale being $10^{-23}$ sec.
Even if we allow extremely dissipative motion of strings, and much
large length scales, the change in quadrupole moment due to strings
can happen in an extremely short time scale (during string
formation, and/or during string decay), compared to the time scale
of rotation, thereby boosting the rate of quadrupole moment change, 
hence power in gravitational wave. For example, even the time scale for heat 
transfer across the core with the speed of sound in the QGP phase 
($\sim 1/\sqrt{3}$) will still give 1000 times smaller time scale
than the fastest spinning time, with a factor of 10$^{9}$ enhancement
in the third time derivative of the quadrupole moment just
due to short time scale \cite{gwave}.

\subsection{Superfluid transition: formation of vortex network}

  Finally, we have also considered superfluid transition occurring inside
a core of size about 3-5 km in the neutron star. As we discussed in Sect.I,
a QCD transition occurring in a core size of few hundred meters will drive
transition to the normal phase (for a pre-existing superfluid phase) in
about 10 times larger radius in the neutron star, i.e. several Km.
Subsequent cooling will lead to superlfuid transition with associated
formation of a dense network of superfluid vortices. We will not necessarily
focus on this particular mechanism of superfluid transition. Our calculations
refer to any superfluid transition happening in a core of several Km size
inside a neutron star. It is important to
note that for a rapidly rotating star, the newly formed network may be
somewhat different from the case of no rotation due to the effect of
rotation biasing the vortex formation (see, ref.\cite{skyrm} for a 
similar biasing). However, in the absence of a detailed model accounting
for this we will simply use the standard simulation of random vortex
network formation during the transition. The free energy density of
transition is taken to be of order 0.1 MeV/$fm^3$ and the vortex
energy per unit length is taken to vary from 1 Mev/fm to about 100 MeV/fm
(ref.\cite{superfluid}). The correlation length for the vortex
formation is taken to be of order 10 fm. Net fractional change in the 
moment of inertia is dominated by the phase change and is of order 
$10^{-6}$ whereas the string induced fractional change in moment of 
inertia is about 4 orders of magnitude smaller, of order $10^{-10}$.
Quadrupole moment and off-diagonal components of moment of inertia
are also found to be of order $10^{-10}$. We note that these numbers
are not far from the values for a glitch (or anti-glitch,
depending on the sign of change of moment of inertia). The transient change
in the moment of inertia decays away when the string system coarsens.
Thus one expects a net rapid change in the spinning rate and restoration
of only few percent of the original value.

\section{conclusions}

 To summarize our results, we have shown that various
phase transitions occurring inside the core of neutron star may
occur on very short time scales (e.g. due to very low nucleation rate
of bubbles inside a large supercritical core), leading to rapid 
changes in the moment of inertia of the star. This directly alters the 
star's rotation rate which  can be detected from pulsar timings. Further, 
density inhomogeneities produced by various phase transitions lead to 
transient change in the MI of the star. 
Such density fluctuations in general lead to  
non-zero off-diagonal components of moment of inertia tensor which will 
cause the wobbling of pulsar, thereby modulating the peak intensity of the 
pulse.  This is a distinguishing and falsifiable signature of our model.
The conventional vortex de-pinning model of glitches 
is not expected to lead to additional wobble as the change in rotation 
caused by de-pinning of vortex clusters remains along the rotation axis. 
We find that moment of inertia can increase or decrease, which gives 
the possibility of accounting for the phenomenon of glitches and anti-glitches
in a unified framework. Development of nonzero value of quadrupole moment
(on a very short time scale) gives the possibility of gravitational
radiation from the star whose core is undergoing a phase transition. We 
emphasize that this is an entirely different way of probing phase transitions 
occurring inside the core of a neutron star, by focusing on minute changes in
one of the most accurately determined quantities in astrophysics, that is
pulsar timings. Though our estimates suffer from the uncertainties of
huge extrapolation involved from the core sizes we are able to simulate,
to the realistic sizes, they strongly indicate that expected changes
in moment of inertia etc. may be well within the range of observations,
and in fact may be able to even account for the phenomena of
glitches and anti-glitches. With much larger simulations, and accounting
for statistical fluctuations of temperature, chemical potential etc. in
the core, more definitive patterns of changes in moment of inertia 
tensor/quadrupole moment etc. may emerge which may carry unique 
signatures of specific phase transitions involved. (For example,
continuous transitions will lead to critical density fluctuations, and
topological defects will induce characterisitic density fluctuations depending
on the specific symmetry breaking pattern.) If that happens then this method 
can provide a rich observational method of probing the physics of strongly
interacting matter in the naturally occurring laboratory, that is 
interiors of neutron star. It will be interesting to see if any 
other astrophysical body, such as white dwarf, can also be probed in a 
similar manner.

\section{ACKNOWLEDGMENTS}
\noindent We thank Ananta P. Mishra, Ranjita Mohapatra, Abhishek Atreya, 
and Shreyansh S. Dave for useful discussions. One 
of the authors, B.Layek did this collaborative work during 
his stay at the Institute of Physics (IoP), Bhubaneswar on sabbatical leave 
from Physics Dept., BITS-Pilani, Pilani. He would like to thank 
the members of HEP group for invitation to IoP and providing all 
facilities and hospitality.

\end{document}